\documentclass[11pt]{article}
\usepackage{a4,amsmath,amssymb,amsthm,amscd,cite,graphicx}
\usepackage{rotating}
\usepackage[table]{xcolor} 
\usepackage{mathtools, booktabs, caption, float, multirow, comment} 
\usepackage{fixmath, geometry} 
\usepackage[hidelinks]{hyperref} 
\usepackage{tikz}
\usetikzlibrary{intersections,calc,matrix,arrows, decorations.markings,shapes}
\usepackage{multirow}

\def\Bbb{\mathbb} \def\BZ{\Bbb Z}  \def\BC{\Bbb C}

\newcommand{\be}{\begin{equation}}
\newcommand{\ee}{\end{equation}}
\newcommand{\bea}{\begin{eqnarray}}
\newcommand{\eea}{\end{eqnarray}}

\catcode`@=11 \@addtoreset{equation}{section} \catcode`@=12

\begin{document}

\begin{titlepage}
\begin{flushright}
\texttt{arXiv:2503.23761  [hep-th]}\\
(v 1.2) August 2025
\end{flushright}
\begin{center}
\textsf{\large Two approaches to the holomorphic modular bootstrap}\\[12pt]
Suresh Govindarajan$^{\dagger}$ and Jagannath Santara$^{*}$ \\
Email: $^\dagger$suresh@physics.iitm.ac.in, $^*$jagannath.santara@physics.iitm.ac.in\\[6pt]
Department of Physics,\\
Indian Institute of Technology Madras\\
Chennai 600036, INDIA
\end{center}
\begin{abstract}
The holomorphic bootstrap attempts to classify rational conformal field theories. The straight ahead approach is hard to implement when the number of characters become large. We combine all characters of an RCFT to form a vector valued modular form with multiplier. Using known results from the theory of vector valued modular forms, given a known RCFT, we obtain new  vector valued modular forms that share the same multiplier as the original RCFT. By taking particular linear combinations of the  new solutions, we look for and find new admissible solutions. In the well-studied two character case, we reproduce all known admissible solutions with Wronskian indices $6$ and $8$. The method is illustrated with examples with up to six characters. The method using vector valued modular forms thus provides a new approach to the holomorphic modular bootstrap.
\end{abstract}
\end{titlepage}
\tableofcontents    

\section{Introduction}

The holomorphic modular bootstrap approach to classifying rational conformal field theories (RCFT) has its origin in the seminal work of Mathur, Mukhi and Sen(MMS) \cite{Mathur:1988na,Mathur:1988gt} and the work of \cite{Naculich:1988xv}. The more recent works of Hamapura-Mukhi\cite{Hamapura:2015} and Chandra-Mukhi\cite{Chandra:2018pjq} has lead to a revival of this approach. In this approach, one treats all characters of an RCFT as linearly independent solutions of a modular linear differential equation (MLDE). The order of the MLDE is the number of characters that appear in the RCFT ignoring multiplicity. Another parameter, the Wronskian index, defined in the sequel, is a second parameter that refines the classification. The holomorphic bootstrap corresponds to looking for solutions of the MLDE whose $q$-series has non-negative integral coefficients. Such solutions are deemed admissible. In general, it does not follow that an admissible solution has an RCFT associated with it. These solutions of an $n$-th order MLDE are modular forms of a level $n$ sub-group of $PSL(2,\BZ)$.

This approach has lead to the classification of all RCFTs with two primaries and central charge less than 25\cite{Mukhi:2022bte}. Further, potential RCFTs with two characters with large Wronskian indices have been obtained in \cite{Das:2023qns}.
It turns out that all known solutions are expressible in terms of the ${}_2F_1$ hypergeometric function. A mathematical approach was carried out by Franc and Mason\cite{Franc:2013} for two-character theories. A generalised coset construction that relates families of two-character theories leads to a CFT understanding of theories obtained from the holomorphic modular bootstrap\cite{Gaberdiel:2016zke}.

The extension of the holomorphic bootstrap to three characters appears implicitly in the work of Franc and Mason\cite{Franc:2015, Franc:2019} where they show that some solutions are expressible in terms of the ${}_3F_2$ hypergeometric function. It turns out that for three characters, all hypergeometric solutions  necessarily have Wronskian index $\ell=0$. It appears that only two character with arbitrary Wronskian index and three character theories with vanishing Wronskian index are the \textit{only} ones that can be expressed in terms of hypergeometric functions. The straightahead approach to the holomorphic modular bootstrap to three characters and arbitrary Wronskian index has been done in a series of papers\cite{Mukhi:2020gnj,Das:2021uvd,Gowdigere:2023xnm}. A scan of known RCFTs obtaining their Wronskian indices was carried out\cite{Das:2020wsi}. This is a useful input to the MLDE approach.  There are some   results for the four character case in \cite{Rayhaun:2023pgc}. A status report dating back to 2019 is given in \cite{Mukhi:2019xjy}. There is an extension of the holmorphic modular bootstrap to  sub-groups of $PSL(2,\mathbb{Z})$ associated with different spin structures on a torus.\cite{Bae:2020xzl,Bae:2021mej}. A flavored refinement of the holomorphic bootstrap has been recently studied in \cite{Pan:2024dod}.

The MLDE approach is hard to carry out as the number of characters exceeds three. One needs to supplement this approach by other approaches. An interesting approach making use of the sub-group $PSL(2,\BZ_n)$ of $PSL(2,\BZ)$ to classify possible values of central charges and values of conformal dimensions after relaxing the condition on non-negativity of the coefficient of the $q$-series\cite{Kaidi:2021ent}. This work provides inputs to the holomorphic modular bootstrap for up to 5 characters.
Another approach is to construct new solutions to the MLDE starting from the characters of an RCFT. For instance, acting on the character of an RCFT by a Hecke operator may lead to another character with a different central charge.\cite{Harvey:2019qzs,Duan:2022ltz}. 

This paper provides another attempt at constructing new admissible characters starting from the characters of a known RCFT. The approach begins with the work of Bantay and Gannon on vector valued modular forms (VVMF)\cite{Bantay:2001ni,Bantay:2005vk,Bantay:2010uy,Gannon:2013jua}. Individual characters of an RCFT are modular forms of a level $n$ subgroup of $PSL(2,\BZ)$, where $n$ is the number of characters in the RCFT. It is useful to think of the $n$ characters as a rank $n$ vector valued modular form with multiplier determined by the modular  $S$ and $T$ matrices of the RCFT. In \cite{Gannon:2013jua}, Gannon obtains a basis of solutions form VVMFs with a given character and weight as a solution to a matrix MLDE. Using the modular theory developed by Gannon and Bantay\cite{Bantay:2010uy,Gannon:2013jua}, we begin the VVMF associated with a given RCFT and construct other solutions that share the same multiplier and hence have the same $S$ and $T$ matrices as the original RCFT. This has the advantage that one can work with larger number of characters that are not approachable using the MLDE approach. In the MLDE approach, the modular $S$ and $T$ matrices are not easily obtained.

The organisation of the paper is as follows. Section 1 gives an introduction to problem being studied. In section 2.1, we provide a brief review to the MMS approach to the holomorphic modular bootstrap setting the notation for future discussion.  In section 2.2, we discuss the VVMF approach to the holomorphic bootstrap. We then show how one can generate new solutions given an RCFT describing the process in detail. In section 2.3, we then provide a method that starts from a $(n,\ell)$ RCFT with central charge $c$ and leads to new VVMFs
that have the same character as the original RCFT. In section 2.4, we study linear combinations of the new solutions with the characters of the original RCFT. Such linear combinations are potentially RCFTs with central charge $c+24$ and Wronskian index $(\ell + 6r)$ for some $r>0$. In section 3, we discuss examples with up to six characters. The examples with five and six characters require a more intricate method to generate new solutions. This is discussed in Section 3.4. We conclude with a brief discussion on future directions and applications of our method.

\section{Two approaches}

\subsection{The MMS approach}

Consider an RCFT  with central charge $c$ and $n$ independent characters, $\chi_i(\tau)$, associated with primaries with conformal dimension $h_i$, $i=0,\ldots,(n-1)$.
As $\tau\rightarrow i\infty$, let
\begin{equation}
\chi_i(\tau) = q^{\alpha_i} \Big(r_0^{(i)} + O(q)\Big)\ ,
\end{equation}
where $\alpha_i = h_i -(c/24)$ with $h_0=0$ and $r_0^{(0)}=1$ and $q=\exp(2\pi i\tau)$.
 In the MMS approach to the classification of RCFT's, the characters are taken to be independent solutions to an $n$-th order modular linear differential equation:
\begin{equation}\label{MDE}
 \Big(\mathcal{D}_0^n  + \sum_{s=1}^n \phi_s(\tau)\ \mathcal{D}_0^{n-s}\Big)\ \chi_i(\tau)=0\ ,
 \end{equation}
 where $\mathcal{D}$ is the modular covariant derivative and $\phi_s(\tau)$ are modular forms of weight $2(n-s)$.
 
The number of zeros of the Wronskian in the finite part of the fundamental domain can be fractional as zeros at the elliptic fixed points,  $\tau=i$ and $\tau=\exp(2\pi i /6)$ are counted with multiplicity $1/2$ and $1/3$ respectively. 
The Wronskian index, $\ell$, is defined to be the six times the number of zeros of the Wronskian at finite values of $\tau$. 
\begin{equation}
\ell = \frac{n(n-1)}{2}-6\sum_{i=0}^{n-1}\, \alpha_i\ . \label{ellformula}
\end{equation}
One has $\ell=0$ or $\ell>1\in \mathbb{Z}$. Thus a zero not located at an elliptic point contributes $6$ to the Wronskian index.

In the MMS approach, one fixes the number of characters and the Wronskian index, i.e., $(n,\ell)$ in our notation. Next, one uses modularity to fix the modular functions that appear in \eqref{MDE} -- this implies that the $\phi_k(\tau)$ are rational functions of $E_4(\tau)$ and $E_6(\tau)$ as these generate the ring of modular forms of $PSL(2,\mathbb{Z})$. Then, \eqref{MDE} is determined in terms of a few undetermined constants -- call them $\mu_1,\mu_2,\ldots$. The goal is to determine the values of the undetermined constants for which all the solutions have non-negative integral coefficients.
\begin{enumerate}
\item One solves \eqref{MDE} using the Frobenius power series method. The exponents $\alpha_i$ are solutions to the indicial equation that appears at leading order. The smallest exponent is identified with $\alpha_0=-c/24$ in unitary RCFTs.
\item The Wronskian index imposes a constraint on the sum of the exponents.
\item Expanding the $q$-series for the  characters as follows
\begin{align}\label{CharExp}
\begin{split}
\chi_0(\tau) &= q^{\alpha_0}(1 + m_1 q + m_2 q^2 + m_3 q^3 + \ldots)  \\
\chi_i(\tau) &= q^{\alpha_i}(r_0^{(i)} + r_1^{(i)} q + r_2^{(i)} q^2 + r_3^{(i)} q^3 + \ldots)
\end{split} 
\end{align}
where the coefficients $m_j$ and $r_j^{(i)}$ are non-negative integers. 
\end{enumerate}
In this method for classifying $(n, \ell)$ RCFT, we need to solve $(n, \ell)$ MLDE
imposing the condition of non-negative integrality on the coefficients $m_j$ and $r_j^{(i)}$. Such solutions are termed as \textit{admissible} solutions.
This method gets hard when we increase the number of characters or the Wronskian index with results known only for theories with up to three characters.

\subsubsection*{Characters as a VVMF} 
 
 Form the $n$-dimensional vector $\mathbb{X}$ by combining the independent solutions to the MLDE.
 \begin{equation}\label{charVVMF}
 \mathbb{X}(\tau)=\begin{pmatrix}
 \chi_0(\tau) \\ \chi_1(\tau) \\ \vdots \\ \chi_{n-1}(\tau)
 \end{pmatrix}\ .
 \end{equation}
It turns out that the RCFT characters are the elements of a vector valued modular form (VVMF) of weight zero for modular group $SL(2, \mathbb{Z})$. A vector valued modular form of weight $w$ and rank $n$ is an $n$-tuple, $\mathbb{X}$, that transforms  under modular group as follows:
\begin{equation}
\mathbb{X}\left(\tfrac{a\tau + b}{c\tau + d}\right) = (c\tau+ d)^w \rho\left(\left(\begin{smallmatrix} a & b\\ c & d \end{smallmatrix}\right)\right) \mathbb{X}(\tau) \label{vvmftran}
\end{equation}  
where, the matrix $\left(\begin{smallmatrix} a & b\\ c & d \end{smallmatrix}\right) \in SL(2, \mathbb{Z})$ and $\rho$, the multiplier, is a representation of the modular group:
\bea
\rho: SL(2, \mathbb{Z}) \to GL(n, \mathbb{C}) \ ,
\eea
satisfying a consistency condition under composition in $SL(2, \mathbb{Z})$.
It is known that this representation is completely characterized by two matrices
\bea
T:=\rho\left(\left(\begin{smallmatrix} 1 & 1\\ 0 & 1\\\end{smallmatrix}\right)\right) ,\, S:= \rho\left(\left(\begin{smallmatrix} 0 & -1\\ 1 & 0\\\end{smallmatrix}\right)\right)\ .
\eea  
and let $U:=ST^{-1}$. 

A representation $(w,\rho)$ is called \textit{admissible} if $\rho(I_2)=e^{-\pi i w} \rho(-I_2)$, where $I_2$ is the $2\times 2$ identity matrix.  This is a necessary  condition for the existence of a VVMF with weight $w$ in Gannon's considerations\cite[see Def. 2.1(a)]{Gannon:2013jua}. 
Then, one has $(e^{\pi iw/2}S)^2=1$   and $(e^{2\pi i w/3} U)^3=1$. When $w=0$, one has $S^2=1$  and $U^3=1$. This will be the case in our considerations.

 We will study this object VVMFs in detail in the next subsection.  
           
 \subsection{The VVMF Approach}    
 
 Gannon\cite{Gannon:2013jua} studied the theory of weakly holomorphic VVMF of $PSL(2,\BZ)$ with weight $w$ and multiplier $\rho$. The weakly holomorphic condition allows for poles at the cusps. It is known that the ring of weakly holomorphic modular forms of $PSL(2,\BZ)$ is generated by 
 $\BC[J,E_4,E_6]$, where $J$ is the Klein-$J$ invariant. Gannon addresses a similar question for VVMFs. Let $\mathcal{M}^{!}_w(\rho)$ denote the space of weakly holomorphic VVMFs with weight $w$ and multiplier $\rho$ and let it have dimension $n$.

Consider the three invariant operators:
\bea
\nabla_{1, w} = \frac{E_4\, E_6}{\Delta}\, \mathcal{D}_w\ ,\quad \nabla_{2, w} = \frac{E_4^2}{\Delta}\, \mathcal{D}_w^2\ ,\quad \nabla_{3, w} = \frac{ E_6}{\Delta}\, \mathcal{D}_w^3 \ . \label{BGOpts}  
\eea
The three operators do not change the weight and multiplier of a modular form on which they act. 
 Let $\mathbb{X}\in \mathcal{M}^{!}_w(\rho)$. Then it is shown in the paper \cite[Prop. 3.2]{Gannon:2013jua} that for admissible representations, one has (this result is also present in the earlier paper \cite{Bantay:2010uy})
 \begin{equation}\label{VVMFring}
 \boxed{
 \mathcal{M}^{!}_w(\rho) = \BC[J,\nabla_{1,w},\nabla_{2,w},\nabla_{3,w}]\ \mathbb{X}\ .
 }
 \end{equation}
Further, $\mathcal{M}^{!}_w(\rho)$, is a $\BC[J]$ module of rank $n=\text{rank}(\rho)$.\cite[Theorem 3.3 (a)]{Gannon:2013jua}.  This implies that given $\mathbb{X}$ with a known admissible representation $(w,\rho)$, one can construct a basis for VVMFs with this weight and multiplier. Further, any VVMF can be written as a linear combination of this basis with coefficients being polynomials in $J$. This has practical implications as we show.

 The main idea of this paper is to use this result to generate \textit{new} solutions using the VVMF $\mathbb{X}$ consisting of $n$ characters of a known RCFT(as in \eqref{charVVMF}). In particular, $\mathbb{X}$ has weight $w=0$ and the $S$ and $T$ matrices are known.  Call these new solutions $\mathbb{Y}_i$ for $i=1,\ldots,(n-1)$. We will find that typically we  find solutions whose $q$-expansion have integral coefficients that are not necessarily non-negative. Solutions with integral coefficients that might be negative are called \textit{quasi-characters}. 

Form the $n\times n$ matrix, $\Xi$, whose columns are $\mathbb{X}(\tau)$ and $\mathbb{Y}_j(\tau)$.  
\bea
\Xi(\tau) = \begin{pmatrix}  \chi_0(\tau) & \chi_0^{(1)}(\tau) & \chi_0^{(2)}(\tau) & \cdots & \chi_0^{(n-1)}(\tau)\\\chi_1(\tau) & \chi_1^{(1)}(\tau) & \chi_1^{(2)}(\tau) & \cdots & \chi_1^{(n-1)}(\tau)\\ \vdots & \vdots & \vdots & \cdots & \vdots\\\chi_{n-1}(\tau) & \chi_{n-1}^{(1)}(\tau) & \chi_{n-1}^{(2)}(\tau) & \cdots & \chi_{n-1}^{(n-1)}(\tau)\\\end{pmatrix}
\eea   
By construction, one has
\bea
\Xi\left(\tfrac{a\tau + b}{c\tau + d}\right) =  \rho\left(\left(\begin{smallmatrix} a & b\\ c & d \end{smallmatrix}\right)\right) \ \Xi(\tau)  \ .
\eea
This implies that all the new solutions have the same $S$ and $T$ matrices as $\mathbb{X}$.
 The $T$ matrix is diagonal and is given by $$T=\text{Diag}(e^{2\pi i\alpha_0},\ldots,e^{2\pi i\alpha_{n-1}})\ .$$ We make the additional assumption that the representation $(w=0,\rho)$ is admissible i.e., $S^2=U^3=I$.\footnote{CPT invariance of an RCFT maps a character to its conjugate character. Thus, in general one can have $S^2=C$ where $C$ is the charge conjugation matrix. Hence, $S^4=I$\cite[see Eq. 10.206]{DiFrancesco:1997nk}. We thus restrict our considerations to RCFT's for which $S^2=I$.}
Let $(a_0,a_1)$ be  the number of eigenvalues of $S$ with eigenvalue $(+1,-1)$ respectively. Similarly, Let $(b_0,b_1,b_2)$ be  the number of eigenvalues of $U$ with eigenvalue $(+1,e^{2\pi i/3},e^{4\pi i/3})$ respectively.

One introduces the matrix of exponents $\Lambda=\text{Diag}(\lambda_0,\ldots,\lambda_{n-1})$ such that $T=\exp(2\pi i \Lambda)$.  There exists a specific choice of the exponents $\lambda_i$ such that a particular map discussed in \cite[Theorem 3.2(b)]{Gannon:2013jua} is invertible. We call this choice admissible. Further, one then has the $q$-expansion
\begin{equation}\label{Ynorm}
\Xi(\tau) = q^\Lambda \left (I_n + \mathcal{Y}\,q+ O(q^2) \right)\ ,
\end{equation}
where $I_n$ is the $n\times n$ identity matrix and $\mathcal{Y}$ is an $n\times n$ matrix. Note that this fixes the normalisation of the VVMFs. Comparing this with \eqref{CharExp}, we see that
\begin{equation}\label{lambdalpharel}
\lambda_0=\alpha_0\quad,\quad \lambda_i = (\alpha_i-1)\text{ for } i\neq 0\ .
\end{equation}
 An application of the Riemann-Roch theorem implies that
\begin{equation}\label{RiRoc1}
\sum\limits_{i=0}^{n-1} \lambda_i =  - \frac{a_1}{2} - \frac{(b_1 + 2\, b_2)}{3}\ . 
\end{equation}  
Comparing the above equation with \eqref{ellformula} and using \eqref{lambdalpharel}, we obtain
\begin{equation}\label{refine}
\boxed{
\ell = \frac{(n-1)(n-12)}{2} + 3 a_1 + 2(b_1+2b_2)\ .
}
\end{equation}
This is an interesting formula as it relates the Wronskian index to the behaviour at the elliptic points.
The numbers $(a_1,b_1,b_2)$ thus provide a \textit{refinement} refinement of the Wronskian index. Let us see how this happens in some examples with low rank.
\begin{enumerate}
    \item For $n=2, \ell=0$, one has $a_1=b_1=1$ and $b_2=0$.
    \item For $n=2, \ell=2$, one has $a_1=b_2=1$ and $b_1=0$.
    \item For $n=3, \ell=0$, one has (i) $a_1=3$ and $b_1=b_2=0$ or (ii) $a_1=b_1=b_2=1$. This is the first instance, where we can see a refinement with two possibilities. Possibility (i) leads to the S-matrix being proportional to the identity matrix. This is equivalent to the tensor product of three single-character theories. Hence, only possibility (ii) can occur.
      \item For $n=3, \ell=3$, we have (i) $a_1=2$ and $b_1=b_2=1$ or (ii) $a_1=0$ and $b_1=0$, $b_2=3$. Possibility (ii) is not considered as it implies that $S$ and $U$ matrices are proportional to identity.
\end{enumerate}
The refinement tracks the behaviour of the solutions at the elliptic points.
From a MLDE viewpoint, this relationship holds only when the MLDE has singularites at the cusp and elliptic fixed points. This is implicitly assumed in our considerations. 
The refinement gives the conjugacy class of the monodromy  matrix about the elliptic fixed points. 

Given a multiplier $\rho$,  the matrix of admissible exponents $\Lambda$ and the $n\times n$ matrix, $\mathcal{Y}$, defined in Eq. \eqref{Ynorm}, Gannon shows that a basis for $\mathcal{M}_w^!(\rho)$ is generated by the solution to the following first-order Matrix MLDE\cite{Gannon:2013jua}:
\begin{equation}\label{MMDE}
\nabla_{1,w}\ \Xi(\tau) = \Xi(\tau)\Big((J(\tau)-984)\ \Lambda_w +\mathcal{Y}_w +[\Lambda_w,\mathcal{Y}_w]\Big)\quad,
\end{equation}
where $\Lambda_w = \Lambda - \frac{w}{12} \mathbf{1}_n$ and $\mathcal{Y}_w = \mathcal{Y} + 2w \mathbf{1}_n$.

\subsection{Generating new solutions from a known solution}

Using the result \eqref{VVMFring} as well as the leading behaviour of the VVMF's $\mathbb{X}$ and $\mathbb{Y}_i$ (for $i=1,\ldots,(n-1)$), we obtain
\be \label{basicnew}
\nabla_a\, \mathbb{X}(\tau) - (\alpha_a\, J(\tau) + \beta_a)\mathbb{X}(\tau) = \sum_{j=1}^{n-1} \gamma_{a,j}\, \mathbb{Y}_j(\tau), 
\ee
The constants $\alpha_a$ and $\beta_a$ are used to cancel the two leading coefficients in  $q$-expansion  of the first row  of the $\nabla_a\, \mathbb{X}(\tau)$. This is consistent with the $\mathbb{Y}_j$ all having $q^{\lambda_0+1}$ as the leading coefficient in the $q$-expansion of the first row.
Recall from Eq. \eqref{Ynorm} that the normalisation of the $\mathbb{Y}_j$ is such that the leading coefficient of the $(j+1)$-th row is one. Then for $j>0$, the constants $\gamma_{a,j}$ are given by the coefficient of $q^{\lambda_j}$  in the $q$-expansion of the $j$-th row  of the $\nabla_a\, \mathbb{X}(\tau)$. This determines up to three independent linear combinations of the other solutions. The precise number is given by the rank of the  matrix $\gamma_{a,j}$. 

To illustrate this idea, we consider the Ising model which is a $(3,0)$ RCFT with $c=\frac12$, $h_0=0$, $h_1=\frac12$ and $h_2=\frac1{16}$. The VVMF of characters is
\be
\mathbb{X} = \left(\begin{array}{l} q^{-\frac{1}{48}}\left(1 + q^2 + q^3 + 2 q^4 + 2 q^5 +\ldots\right)\\ 
q^{\frac{23}{48}}\left(1 + q + q^2 + q^3 + 2 q^4 + 2 q^5 + \ldots\right)\\ q^{\frac{1}{24}}\left(1 + q + q^2 + 2 q^3 + 2 q^4 + 3 q^5 +\ldots\right)\end{array}\right)
\ee
Applying $\nabla_1$ and $\nabla_2$, we obtain
\begin{align*}
\nabla_1\mathbb{X} -\left(-\tfrac1{48} J +\tfrac{41}{2}\right) \mathbb{X} &= \tfrac{1}{2} \mathbb{Y}_1 + \tfrac{1}{16}\mathbb{Y}_2\ .\\
\nabla_2\mathbb{X} -\left(\tfrac{1}{256} J -\tfrac{49}{48}\right) \mathbb{X} &=\tfrac{7}{48} \mathbb{Y}_1 -\tfrac{7}{768}\mathbb{Y}_2\ .
\end{align*}
This gives us two other VVMFs: 
\begin{align*} 
\mathbb{Y}_1 &= \left(\begin{array}{l} q^{ \frac{47}{48}}\left(2325 + 60630 q + 811950 q^2 + 7502125 q^3 + 54345125 q^4 + 
 329961125 q^5 +\ldots\right)\\ q^{-\frac{25}{48}}\left(1 + 275 q + 13250 q^2 + 235500 q^3 + 2558550 q^4 + 20713510 q^5 + \ldots\right)\\ q^{\frac{1}{24}}\left(-25 - 4121 q - 102425 q^2 - 1331250 q^3 - 12083250 q^4 - 86425675 q^5 +\ldots\right)\end{array}\right)  \\
 \mathbb{Y}_2 &= \left(\begin{array}{l} q^{ \frac{47}{48}}\left(94208 + 9515008 q + 356765696 q^2 + 7853461504 q^3 + 
 122161491968 q^4   +\ldots\right)\\ q^{ \frac{23}{48}}\left(-4096 - 1130496 q - 63401984 q^2 - 1763102720 q^3 - 32112021504 q^4  + \ldots\right)\\ q^{-\frac{23}{24}}\left(1 - 23 q + 253 q^2 - 1794 q^3 + 9384 q^4   + \ldots\right) \end{array}\right) 
 \end{align*}     
The solution $\mathbb{Y}_1$ becomes an admissible character when we change the sign of the third row. The second row of $\mathbb{Y}_1$ is the identity character and thus $c=\frac{25}2$, $h_1=\frac32$ and $h_2=\frac9{16}$.  Let $\mathbb{Y}_1'$ denote the VVMF obtained from $\mathbb{Y}_1$ by changing the sign of the third row and exchanging rows 1 and 2. Changing the sign of the third row changes the $S$-matrix. However, the exchange of rows 1 and 2, makes the $S$-matrix of $\mathbb{Y}_1'$ identical to that of $\mathbb{Y}_1$\footnote{The S-matrix for the Ising model is given by
\[
S=\begin{pmatrix}\frac{1}{2} &\frac{1}{2} &\frac{1}{\sqrt{2}}\\
    \frac{1}{2} &\frac{1}{2} & -\frac{1}{\sqrt{2}}\\
     \frac{1}{\sqrt{2}} &-\frac{1}{\sqrt{2}} & 0
     \end{pmatrix}\ .
\]}. Note that this does not happen generically and is special to this example.
This solution was obtained using the MLDE method in \cite{Das:2021uvd}. 
This CFT is the GHM coset dual of the $B_{11,1}$ CFT (the $B_{11}$ Kac-Moody Lie algebra at level one) with $c=\tfrac{23}2$\cite{Gaberdiel:2016zke}. Let $\mathbb{Z}$ denote the VVMF of the $B_{11,1}$ CFT. Explicitly, one has
\begin{equation}
\mathbb{Z}= \left(\begin{array}{l}
q^{-\frac{23}{48}}(1+253 q + 9384 q^2+ \ldots )\\
q^{\frac1{48}}(23 + 1794 q + 39491 q^2 + \ldots )\\
q^{\frac{23}{24}}(2048 + 47104 q + 565248 q^2 +\ldots)
\end{array}\right)\ .
\end{equation}
Then, the GHM coset condition in this case is
\[
\mathbb{Z}^T\cdot \mathbb{Y}_1' = J-216 .
\]
This is a special case of the following connection. Let $r\in[0,11]$. Then
\[
\boxed{B_{r,1} \text{ CFT}}\quad \longleftrightarrow\quad \boxed{\text{GHM dual of }B_{11-r,1} \text{ CFT}}\ ,
\]
where $B_{0,1}$ refers to the Ising model\cite{ongoing}.

For $n\leq 4$, this can completely determine the $\mathbb{Y}_i$ when  $\gamma_{a,j}$ has maximal rank.  We will discuss multiple examples in the next section.
This method does not give all solutions when $n>4$ or when $\gamma$ doesn't have maximal rank. In such cases, we need to take multiple derivatives until we obtain $(n-1)$ linear independent combinations. We will illustrate this method with examples with $n=5,6$.

\subsection{Going from quasi-characters to admissible ones}

The new solutions $\mathbb{Y}_i$ may have integral coefficients that may not be non-negative. As mentioned earlier, we then call them quasi-characters. An important fact is that the $\mathbb{Y}_i$ have the same multiplier as $\mathbb{X}$. Thus taking linear combinations of $\mathbb{X}$ with  the $\mathbb{Y}_i$, with coefficients that are polynomials in the $J$ function, does not affect the modular properties. This may lead to admissible solutions\cite{Chandra:2018pjq}. We do not modify the $\mathbb{Y}_i$ by changing signs of one of the rows or exchanging rows as we did in the Ising model example. This generically changes the multiplier and we can no longer take linear combinations. 

\subsubsection*{Taking direct linear combinations}

Consider the linear combinations
\begin{equation}
\mathbb{U}_r= \mathbb{X} + \sum_{i=1}^r b_i \mathbb{Y}_i \ , \label{Udef}
\end{equation}
where $r\in[1,n-1]$ and $b_i>0$ are some positive integral constants\footnote{For simplicity, we pick the first $r$ quasi-characters when any $r$ of the $(n-1)$ quasi-characters may be chosen.}. Let us assume that for some ranges of these positive integers, we obtain admissible solutions. As the $\mathbb{Y}_i$ are typically quasi-characters, we will find upper bounds on the constants. In such cases, we obtain an admissible solution with the \textit{same} central charge but with Wronskian index $(\ell + 6r)$.

In the example of the Ising model,  there is no solution where the $U_r$ are admissible. However, we will see other examples where this is not the case.

\subsubsection*{Other combinations}

Consider the VVMF $J(\tau)\, \mathbb{X}$ -- this is the CFT obtained by tensoring the original $(n,\ell)$ CFT by a single character CFT and is a $(n,\ell+6n)$ CFT with the same multiplier as the original CFT. For some $r\in [1,n-1]$, consider the combination
\begin{equation}
\mathbb{W}_r =(J(\tau) + b)\ \mathbb{X} - \sum_{i=1}^{r} r_0^{(i)}\ \mathbb{Y}_i \ ,
\end{equation}
where $b$ is some constant and the $r_0^{(i)}$ are as defined in \eqref{CharExp}. This choice minimises the value of the Wronskian index of $\mathbb{W}_r$ to $(\ell+6(n-r))$. It  leads to the exponents of $\mathbb{W}_r$ being 
\[(\alpha_0-1,\alpha_1,\ldots,\alpha_r,(\alpha_{r+1}-1),\ldots,(\alpha_{n-1}-1)) \ .
\]
 However, it can happen that for a particular value of $b$, the index can be made smaller by $6$. 
Generically, the central charge is $(c+24)$, and weights are 
\[
(h_0=0,(h_1+1),\ldots,(h_r+1),h_{r+1},\ldots,h_{n-1})\ .
\]
Hence, if for some values of $b$, $\mathbb{W}_r$ is admissible, then we obtain a CFT with $(n,\ell+ 6(n-r))$ with the same multiplier as the original CFT. In this fashion, we obtain a series of potential CFTs 
\[
(n,\ell) \rightarrow (n,\ell+6) \rightarrow (n,\ell+12) \rightarrow \cdots\rightarrow (n,\ell + 6n)
\]
We illustrate the method using the Ising model example. In this model,
\[
\mathbb{W}_2 =(J(\tau) + b)\ \mathbb{X} - \mathbb{Y}_1 -\mathbb{Y}_2
\]
is an admissible character if $b\geq-744$. 
\begin{align*} 
\mathbb{W}_2  = \left(\begin{array}{l} q^{ -\frac{49}{48}}\left(1 + (744 + b) q + 
 100352 q^2 + (11918867 + b) q^3 + (506919954 + 
    b) q^4    +\ldots\right)\\ q^{ \frac{23}{48}}\left((4566 + b) + (1314875 + b) q + (84857873 + b) q^2 + (2646535530 + 
    b) q^3    + \ldots\right)\\ q^{\frac{1}{24}}\left((793 + b) + (201497 + b) q + (21795609 + b) q^2 + (887313970 + 
    2 b) q^3    +\ldots\right)\end{array}\right)  
 \end{align*}
It has central charge $c=(\frac{1}2+24)$, $(h_1=\tfrac12 + 1)$ and $h_2 =(\tfrac1{16}+1)$. It is an admissible character with Wronskian index $\ell=6$.
For $r=1$, we have two possibilities
\[
\mathbb{W}_1 =(J(\tau) + b)\ \mathbb{X} - \mathbb{Y}_1\ \text{and}\ 
\mathbb{W}_1' =(J(\tau) + b)\ \mathbb{X} - \mathbb{Y}_2 \ .
\]
Explicitly, one has
\begin{equation*}
\mathbb{W}_1=
\left(\begin{array}{l}
q^{ -\frac{49}{48}}(1+(b+744)q +194560 q^2+(b+21433875) q^3+\ldots)\\
q^{\frac{23}{48}}((b+470)+(b+184379) q+(b+21455889) q^2+(b+883432810)
   q^3+\ldots) \\
 q^{ -\frac{49}{48}}(  1+(b+770)q+(b+201750) q^2+(b+21793815) q^3+\ldots)
 \end{array}\right)  
\end{equation*}
For $b\geq -470$, $\mathbb{W}_1$ an admissible character. For $b=-470$, the solution has Wronskian index $6$ and for $b>-470$, the solution has Wronskian index $12$. 
For $b\geq -744$, $\mathbb{W}_1'$ is admissible and has Wronskian index $12$. 
\begin{equation*}
\mathbb{W}_1'=
\left(\begin{array}{l}
q^{ -\frac{49}{48}}(1+(b+744)q+102677 q^2+(b+11979497) q^3+\ldots)\\
q^{ -\frac{25}{48}}(1+(b+4841)q+(b+1328125) q^2+(b+85093373) q^3\ldots)\\
  q^{\frac1{24}}( (b+768)+(b+197376) q+(b+21693184) q^2+(2 b+885982720) q^3+\ldots)
    \end{array}\right)
\end{equation*}
The Ising model example illustrates how one can generate new admissible solutions.

\section{Examples}

In this section, we discuss explicit examples in increasing order of complexity -- this is determined by the number of characters. For more than four characters, the method required is more intricate and we discuss examples involving five and six characters. 

\subsection{Two character examples}

For the two-character case, one has
\begin{equation}
\nabla_1\mathbb{X}(\tau)= (\alpha J(\tau)+\beta) \, \mathbb{X}(\tau)+ \gamma\, \mathbb{Y}(\tau)\ .
\end{equation}
This determines the new solution $\mathbb{Y}$. First, consider the combination
\begin{equation}
\mathbb{U}= \mathbb{X}+ b\, \mathbb{Y}\ .
\end{equation}
For the $(2,0)$ MMS RCFT's we find no new admissible solutions of this form.\footnote{An argument due to Sunil Mukhi that we reproduce here explains this result. One can prove that such VVMFs don't exist, using quasi-characters, as follows. At $c=1$ there is one set of $(2,0)$ admissible characters, the MMS set. It is known (Kaneko et al.\cite{Kaneko:2013uga}) that this set is complete. Now there are quasi-characters (QC) with the same $S$,$T$ matrices having $c=-23$ as well as $c=25$. However if we combine $c=1$ and $c=-23$, there are no admissible solutions. This is because $|-23|>1$, so the type II asymptotic behaviour of $c=-23$ dominates (in the non-identity character) by Cardy-type arguments. This is why there are no $(2,6)$ VVMF with the same c as the MMS solutions.} However, for the $(2,2)$ RCFT's that are  coset (GHM) duals of the $(2,0)$ theories\cite{Gaberdiel:2016zke}, we obtain admissible  solutions for $0<b\leq b_{\text{max}}$. The details of this is shown in Table \ref{tab0}. This agrees with the solutions  obtained in \cite{Mukhi:2022bte,Das:2023qns}. In particular, this generates the first half of the $(2,8)$ admissible characters listed in Eq. (3.61) of \cite{Das:2023qns}. 

\setlength{\arrayrulewidth}{0.3mm} 
\setlength{\tabcolsep}{9pt}
\renewcommand{\arraystretch}{1.3}
\definecolor{mColor1}{rgb}{0.9,0.9,0.9}  
\definecolor{mColor2}{gray}{0.8}
\definecolor{mColor3}{gray}{1.0}

\begin{table}[H]  
\centering
\aboverulesep = 0pt
\belowrulesep = 0pt
\begin{tabular}{c||cc|ccc}  
\toprule
 & \multicolumn{2}{c|}{$(2, 2)$ Sol.}      & \multicolumn{3}{c}{$(2, 8)$ Admissible Sol.}   \\  
\cmidrule(lr){2-3} 
\cmidrule(lr){4-6}          
\multirow{-2}{11pt}{Sl. No.}  & $c$ & $h$  & $c$ & $h$ & $b_{\rm max}$\\  
\midrule
1. & $\frac{82}{5}$ & $\frac{6}{5}$ &  $\frac{82}{5}$ & $\frac{1}{5}$ & $2$  \\ \rowcolor{mColor2}
2. & $17$ & $\frac{5}{4}$  &  $17$ & $\frac{1}{4}$ & $5$  \\  
3. & $18$ & $\frac{4}{3}$  &  $18$ & $\frac{1}{3}$ & $9$  \\\rowcolor{mColor2}   
4. & $\frac{94}{5}$ & $\frac{7}{5}$  &  $\frac{94}{5}$ & $\frac{2}{5}$ & $26$  \\ 
5. & $20$ & $\frac{3}{2}$  &  $20$ & $\frac{1}{2}$ & $37$  \\\rowcolor{mColor2}
6. & $\frac{106}{5}$ & $\frac{8}{5}$  &  $\frac{106}{5}$ & $\frac{3}{5}$ & $155$  \\
7. & $22$ & $\frac{5}{3}$ &  $22$ & $\frac{2}{3}$ & $190$  \\\rowcolor{mColor2}
8. & $23$ & $\frac{7}{4}$  &  $23$ & $\frac{3}{4}$ & $498$  \\
9. & $\frac{118}{5}$ & $\frac{9}{5}$  &  $\frac{118}{5}$ & $\frac{4}{5}$ & $591$  \\
\bottomrule
\end{tabular}
\caption{$(2, 8)$ admissible solutions using $(2, 2)$ solutions with central charge unchanged.}
\label{tab0}
\end{table}
\noindent Next consider the combination
\begin{equation}
\mathbb{W}(\tau) = (J(\tau)+b)\,\mathbb{X}(\tau)- r_0\, \mathbb{Y}(\tau)\ .
\end{equation}
Table \ref{tab1} lists the values of $(\alpha,\beta,\gamma)$ for all the $(2,0)$ RCFTs obtained by MMS. Further, the last column lists the minimum value of $b$ for which $\mathbb{W}$ is admissible. We have checked this for the first fifty terms and hence the positivity is only conjectural. These $(2,6)$ examples have been shown to be realised as RCFTs in \cite{Chandra:2018ezv}.

\setlength{\arrayrulewidth}{0.3mm} 
\setlength{\tabcolsep}{9pt}
\renewcommand{\arraystretch}{1.3}
\definecolor{mColor1}{rgb}{0.9,0.9,0.9}  
\definecolor{mColor2}{gray}{0.8}
\definecolor{mColor3}{gray}{1.0}
\begin{table}[H]  
\centering
\aboverulesep = 0pt
\belowrulesep = 0pt
\begin{tabular}{c||cc|ccc|ccc}  
\toprule
 & \multicolumn{2}{c|}{MMS Sol.}   &   &   &   & \multicolumn{3}{c}{$(2, 6)$ Admissible Sol.}  \\  
\cmidrule(lr){2-3} 
\cmidrule(lr){7-9}          
\multirow{-2}{5pt}{S. No.}  & $c$ & $h$ & \multirow{-2}{5pt}{$\alpha$} &\multirow{-2}{5pt}{$\beta$} &\multirow{-2}{5pt}{$\gamma$} & $c$ & $h$ &$b_{\text{min}}$\\  
\midrule
1. & $\frac{2}{5}$ & $\frac{1}{5}$ & $-\frac{1}{60}$& $\frac{87}{5}$ & $\frac{1}{5}$ &  $\frac{122}{5}$ & $\frac{6}{5}$ & $-745$ \\ \rowcolor{mColor2}
2. & $1$ & $\frac{1}{4}$ & $-\frac{1}{24}$& $44$ & $\frac{1}{2}$ &  $25$ & $\frac{5}{4}$ & $-747$ \\  
3. & $2$ & $\frac{1}{3}$ & $-\frac{1}{12}$& $90$ & $1$ &  $26$ & $\frac{4}{3}$ & $-752$ \\\rowcolor{mColor2}   
4. & $\frac{14}{5}$ & $\frac{2}{5}$ & $-\frac{7}{60}$& $\frac{644}{5}$ & $\frac{14}{5}$ &  $\frac{134}{5}$ & $\frac{7}{5}$ & $-758$ \\ 
5. & $4$ & $\frac{1}{2}$ & $-\frac{1}{6}$& $192$ & $4$ &  $28$ & $\frac{3}{2}$ & $-772$ \\\rowcolor{mColor2}
6. & $\frac{26}{5}$ & $\frac{3}{5}$ & $-\frac{13}{60}$& $\frac{1326}{5}$ & $\frac{78}{5}$ &  $\frac{146}{5}$ & $\frac{8}{5}$ & $-796$ \\
7. & $6$ & $\frac{2}{3}$ & $-\frac{1}{4}$& $324$ & $18$ &  $30$ & $\frac{5}{3}$ & $-822$  \\\rowcolor{mColor2}
8. & $7$ & $\frac{3}{4}$ & $-\frac{7}{24}$& $420$ & $42$ &  $31$ & $\frac{7}{4}$ & $-877$  \\
9. & $\frac{38}{5}$ & $\frac{4}{5}$ & $-\frac{19}{60}$& $\frac{2508}{5}$ & $\frac{228}{5}$ &  $\frac{158}{5}$ & $\frac{9}{5}$ & $-934$  \\
\bottomrule
\end{tabular}
\caption{$(2, 6)$ admissible solutions using $(2, 0)$ solutions.}
\label{tab1}
\end{table}

Similarly, we can generate $(2,8)$ admissible characters starting from the known $(2,2)$ admissible characters. Table \ref{tab2} lists the $(2,8)$ admissible characters that we obtain for $b>b_\text{min}$. This generates the second half of the $(2,8)$ admissible characters listed in Eq. (3.61) of \cite{Das:2023qns}. 
\begin{table}[H]  
\centering
\aboverulesep = 0pt
\belowrulesep = 0pt
\begin{tabular}{c||cc|ccc|ccc}  
\toprule
 & \multicolumn{2}{c|}{(2,2) Sol.}   &   &   &   & \multicolumn{3}{c}{$(2, 8)$ Admissible Sol.}  \\  
\cmidrule(lr){2-3} 
\cmidrule(lr){7-9}          
\multirow{-2}{5pt}{S. No.}  & $c$ & $h$ & \multirow{-2}{5pt}{$\alpha$} &\multirow{-2}{5pt}{$\beta$} &\multirow{-2}{5pt}{$\gamma$} & $c$ & $h$ &$b_{\text{min}}$\\  
\midrule
1. & $\frac{82}{5}$ & $\frac{6}{5}$ & $-\frac{41}{60}$& $\frac{5412}{5}$ & $\frac{5412}{5}$ &  $\frac{202}{5}$ & $\frac{11}{5}$ & $-1154$ \\ \rowcolor{mColor2}
2. & $17$ & $\frac{5}{4}$ & $-\frac{17}{24}$& $1020$ & $2040$ &  $41$ & $\frac{9}{4}$ & $-1067$ \\  
3. & $18$ & $\frac{4}{3}$ & $-\frac{3}{4}$& $972$ & $2916$ &  $42$ & $\frac{7}{3}$ & $-978$ \\\rowcolor{mColor2}   
4. & $\frac{94}{5}$ & $\frac{7}{5}$ & $-\frac{47}{60}$& $\frac{4794}{5}$ &  $\frac{33558}{5}$ & $\frac{214}{5}$ & $\frac{12}{5}$  & $-932$ \\ 
5. & $20$ & $\frac{3}{2}$ & $-\frac{5}{6}$& $960$ & $\frac{15}2$ &  $44$ & $\frac{5}{2}$ & $-884$ \\\rowcolor{mColor2}
6. & $\frac{106}{5}$ & $\frac{8}{5}$ & $-\frac{53}{60}$& $\frac{4876}{5}$ & $\frac{126776}{5}$ &  $\frac{226}{5}$ & $\frac{13}{5}$ & $-850$ \\
7. & $22$ & $\frac{5}{3}$ & $-\frac{11}{12}$& $990$ & $\frac{110}3$ &  $46$ & $\frac{8}{3}$ & $-832$  \\\rowcolor{mColor2}
8. & $23$ & $\frac{7}{4}$ & $-\frac{23}{24}$& $1012$ & $56672$ &  $47$ & $\frac{11}{4}$ & $-813$  \\
9. & $\frac{118}{5}$ & $\frac{9}{5}$ & $-\frac{59}{60}$& $\frac{5133}{5}$ &$\frac{292581}{5}$ &   $\frac{238}{5}$ & $\frac{14}{5}$ & $-803$  \\
\bottomrule
\end{tabular}
\caption{$(2, 8)$ admissible solutions using $(2, 2)$ solutions.}
\label{tab2}
\end{table}

\subsection{A $(3,3)$ example}

The three characters of the potential RCFT with $(c=\frac{48}{7}, h_1 = \frac{1}{7}, h_2 = \frac{5}{7})$ theory are given by
\bea
\mathbb{X} = \left(\begin{array}{l} q^{-\frac{2}{7}}\left(1 + 78 q + 784 q^2 + 5271 q^3 + 26558 q^4 + 113756 q^5 +\ldots\right)\\ 
q^{-\frac{1}{7}}\,\left(1 + 133 q + 1618 q^2 + 11024 q^3 + 56532 q^4 + 240968 q^5 + \ldots\right)\\ 
q^{\frac{3}{7}}\, \left(55 + 890 q + 6720 q^2 + 37344 q^3 + 168077 q^4+\ldots\right)\end{array}\right) 
\eea
This was obtained using the MMS approach in\cite{Gowdigere:2023xnm}. This is solution S-3 (see Appendix A) in their  family of admissible solutions with Wronskian index $3$.

Using this solution we find two new solutions as follows
\begin{align}
&\nabla_1 \mathbb{X} -\left(-\tfrac27 J + \tfrac{2514}7\right)\,\mathbb{X}= \tfrac17 \mathbb{Y}_1 +\tfrac{275}7 \mathbb{Y}_2 \\
&\nabla_2 \mathbb{X} -\left(\tfrac{19}{147} J - \tfrac{575}{49}\right)\,\mathbb{X}= -\tfrac{25}{294} \mathbb{Y}_1 -\tfrac{275}{294} \mathbb{Y}_2
\end{align}
where
\be 
\mathbb{Y}_1 = \left(\begin{array}{l} q^{ \frac{5}{7}}\,\left(45954  + 12285036 q  + 822051533 q^2 + 27921229044 q^3 + 620215644852 q^4 +\ldots\right)\\ q^{-\frac{8}{7}}\left(1  + 3q - 147996 q^2 - 29612018 q^3 - 1753689468 q^4 - 55439260596 q^5  + \ldots\right)\\ q^{\frac{3}{7}}\,\left(2925   + 1684020 q  + 149623344 q^2 + 5954910507 q^3 + 146931310551 q^4 +\ldots\right)\end{array}\right) \label{33Gannonyan1} 
 \ee
 \be
\mathbb{Y}_2 = \left(\begin{array}{l} q^{ \frac{5}{7}}\,\left(1702  + 74037 q  + 1309023 q^2 + 14808436 q^3 + 126021372 q^4+\ldots\right)\\ q^{- \frac{1}{7}}\, 
\left(1  + 1380 q + 51498 q^2 + 841526 q^3 + 9082404 q^4 + 74801592 q^5 + \ldots\right)\\ q^{-\frac{4}{7}}\left(1  - 321 q  - 22860 q^2 - 488656 q^3 - 6142032 q^4 - 56105667 q^5 +\ldots\right)\end{array}\right) \label{33Gannonyan2} 
\ee   

Consider  solutions of the form $\mathbb{U}^{(i)}_1 = \mathbb{X}+b_i \mathbb{Y}_i$ for $i=1,2$.  There are no positive integral values of $b$ where this solution is admissible. Permitting fractions less than one also doesn't seem to help. A similar conclusion appears for $\mathbb{U}_2= \mathbb{X}+b_1 \mathbb{Y}_1+ b_2\mathbb{Y}_2$. So we find no admissible solutions in this fashion.

Consider the new solution $\mathbb{W}_2 = (J+b)\mathbb{X}-\mathbb{Y}_1 - 55 \mathbb{Y}_2$. 
\begin{equation}
\mathbb{W}_2=\left(\begin{array}{l}q^{-\frac97}(1+(b+822)q+(78 b+116136) q^{2}+(784 b+21082208) q^{3}
 +\cdots)\\
   q^{-\frac17}((b+819))+(133 b+369550) q+(1618 b+75673776) q^2+\cdots)\\
    q^{\frac37}(55 (b+1028) +(890 b+11070780) q+\cdots)
   \end{array}\right)
   \end{equation}
For $b=-819$, we obtain an admissible solution with Wronskian index $\ell=3$. This solution was labelled S-14 in the $(3,3)$ solutions obtained via the MLDE method in \cite[see Appendix A]{Gowdigere:2023xnm}\footnote{A similar construction starting from the $c=8$ solution labelled S-6 leads to the $c=32$ solution labelled S-15 in the same paper. We do not provide the details here.}.
For $b>-819$, we obtain admissible solutions with $\ell=9$. 

Next considering $\mathbb{W}_1 = (J+b)\mathbb{X}-\mathbb{Y}_1 $, we obtain admissible solutions with $\ell=15$ for $b\geq -707$. Similarly $\mathbb{W}_1' = (J+b)\mathbb{X} - 55 \mathbb{Y}_2$ leads to admissible characters with $\ell=15$ for $b\geq-822$. 

\subsection{A four-character example}

The three-state Potts model is contained in Virasoro minimal model $\mathcal{M}(6, 5)$ with central charge $c=\frac{4}{5}$ and the conformal dimensions are $\{0, \frac{2}{5}, \frac{7}{5}, \frac{2}{3}, 3, \frac{1}{15}\}$. The six characters are given by
\begin{align} 
 \mathbb{X} (\tau) = \left(\begin{array}{l} q^{- \frac{1}{30}}\left(1 + q^2 + q^3 + 2 q^4 + 2 q^5 + 4 q^6+\ldots\right)\\ q^{  \frac{11}{30}} \left(1 + q + q^2 + 2 q^3 + 3 q^4 + 4 q^5+ \ldots\right)\\ q^{ \frac{41}{30}} \left(1 + q + 2 q^2 + 2 q^3 + 4 q^4 + 5 q^5+\ldots\right) \\q^{ \frac{19}{30}} \left(1 + q + 2 q^2 + 2 q^3 + 4 q^4 + 5 q^5+\ldots\right)\\q^{ \frac{89}{30}} \left(1 + q + 2 q^2 + 3 q^3 + 4 q^4 + 5 q^5+\ldots\right)\\q^{ \frac{1}{30}} \left(1 + q + 2 q^2 + 3 q^3 + 5 q^4 + 7 q^5+\ldots\right)\end{array}\right)     
\end{align} 
We can consider it is a four-character theory where the characters are
\bea
\mathbb{X}(\tau) = \left(\chi_{0} + \chi_{3},\, \chi_{\frac{2}{5} }+ \chi_{\frac{7}{5}},\, \chi_{\frac{2}{3}},\, \chi_{\frac{1}{15}}\right)^T
\eea 
The action of $\nabla_i$, $i=1,2,3$,  on $\mathbb{X}$ leads to three new solutions.
\begin{equation}
\begin{split} 
 \mathbb{Y}_1(\tau) &= \left(\begin{array}{l} q^{ \frac{29}{30}}\left(5681 + 223744 q + 4053612 q^2+\ldots\right)\\ q^{ -\frac{19}{30}} \left(1 - 152 q - 17556 q^2 - 469832 q^3+ \ldots\right)\\ q^{\frac{19}{30}} \left(1196 + 73853 q + 1634380 q^2+\ldots\right) \\q^{\frac{1}{30}} \left(-19 - 4636 q - 170867 q^2+\ldots\right)\end{array}\right), \\
\mathbb{Y}_2(\tau) &= \left(\begin{array}{l} q^{ \frac{29}{30}}\left(594 + 8316 q + 70686 q^2+\ldots\right)\\ q^{  \frac{11}{30}} \left(108 + 2970 q + 33264 q^2+ \ldots\right)\\ q^{-\frac{11}{30}} \left(1 - 88 q - 1881 q^2 - 17996 q^3+\ldots\right) \\q^{\frac{1}{30}} \left(-11 - 572 q - 7876 q^2+\ldots\right)\end{array}\right) \nonumber  
\\
 \mathbb{Y}_3(\tau) &= \left(\begin{array}{l} q^{ \frac{29}{30}}\left(129168 + 13578786 q + 522405072 q^2+\ldots\right)\\ q^{ \frac{11}{30}} \left(-1566 - 613872 q - 39297204 q^2+ \ldots\right)\\ q^{\frac{19}{30}} \left(-8671 - 1675504 q - 83293626 q^2+\ldots\right) \\q^{-\frac{29}{30}} \left(1 + 57478 q^2 + 5477520 q^3+\ldots\right)\end{array}\right) 
  \label{3Pottsothersols}  \nonumber\\ 
\end{split}
\end{equation}
The solution $\mathbb{Y}_3$ corresponds to admissible solution after switching signs on two of the entries. We construct new admissible characters using the newly generated solutions i.e., $\mathbb{Y}_1$, $\mathbb{Y}_2$ and $\mathbb{Y}_3$.

Consider solutions of the form $\mathbb{U}^{(i)}_1 = \mathbb{X}+b_i \mathbb{Y}_i$ for $i=1,2,3$ and similarly for $\mathbb{U}_2$ and $\mathbb{U}_3$. We find no admissible solutions from this construction. However, the second approach which involves multiplying $\mathbb{X}$ by the $J$ function leads to new admissible solutions. The solutions are listed in Table \ref{tab3}.

\begin{table}[h] \centering
\begin{tabular}{c|c|c}
\textbf{Admissible Character} & \textbf{Condition} & $\ell$ \\ \hline
$\mathbb{W}_3 = (J+b)\mathbb{X}-\sum_{i=1}^3 \mathbb{Y}_i$ & $b\geq -744$  & $6$ \\ \rowcolor{mColor2}
$\mathbb{W}_2 = (J+b)\mathbb{X}- \mathbb{Y}_1-\mathbb{Y}_2$ & $b= 363$  & $6$ \\  \hline
$\mathbb{W}_2 = (J+b)\mathbb{X}- \mathbb{Y}_1-\mathbb{Y}_2$ & $b> 363$  & $12$ \\
$\mathbb{W}_2' = (J+b)\mathbb{X}- \mathbb{Y}_1-\mathbb{Y}_3$ & $b\geq-744$  & $12$ \\
$\mathbb{W}_2'' = (J+b)\mathbb{X}- \mathbb{Y}_2-\mathbb{Y}_3$ & $b\geq-744$  & $12$ \\ \hline
$\mathbb{W}_1 = (J+b)\mathbb{X}- \mathbb{Y}_1$ & $b\geq451$  & $18$ \\
$\mathbb{W}_1' = (J+b)\mathbb{X}- \mathbb{Y}_2$ & $b\geq-638$  & $18$ \\
$\mathbb{W}_1''' = (J+b)\mathbb{X}- \mathbb{Y}_3$ & $b\geq-744$  & $18$ \\ \hline

\end{tabular}
\caption{Admissible characters with $c=\frac{124}5$ obtained using the new solutions}\label{tab3}
\end{table}

\subsection{Going beyond four characters}

We will need to find additional solutions beyond those obtained from Eq. \eqref{basicnew}. These are determined by using the result mentioned in Eq. \eqref{VVMFring}. This would lead us to consider combinations such as $\nabla_1\nabla_3 \mathbb{X}$. An alternate method due to Bantay discusses particular combinations that are
best suited for our computations\cite{Bantay:2010uy}. He defines invariant operators that map $\mathcal{M}_w(\rho)$ to itself. 
\begin{equation}
\nabla_4 = \frac{E_4 }{\Delta}\ \mathcal{D}^4\quad, \quad \nabla_5 = \frac{E_4^2 E_6}{\Delta^2}\ \mathcal{D}^5\quad.
\end{equation}
We find the following useful relation
\begin{equation}
\nabla_4 \mathbb{X} - (\alpha_4 J + \beta_4) \mathbb{X} = \sum_{i=1}^{n-1}\gamma_{4,i}\, \mathbb{Y}_i \ ,
\end{equation}
where we choose the constants $(\alpha_4,\beta_4)$ by matching the leading terms in the first row. These are the coefficient of $q^{\alpha_0-1}$ and $q^{\alpha_0}$. We can determine $\gamma_{4,i}$ by studying the leading term in other rows.
Once these are chosen,the right hand side of the above equation is a known linear combination of the unknown solutions exactly as in Eq. \eqref{basicnew}. This is sufficient when one has five characters and the matrix 
$\gamma_{a,j}$ for $a,j\in[1,4]$ is invertible. One has
\[
\mathbb{Y}_i= \sum_{a=1}^4 (\gamma^{-1})_{i,a} \left(\nabla_a \mathbb{X} -(\alpha_a J +\beta_a)\mathbb{X}\right)\ .
\] 

When one has more than five characters and $\gamma_{a,i}$ has rank four, let $M_{ab}$ be such that
\[
\sum_{b=1}^4 M_{ab}\, \gamma_{b,i} =\left\{\begin{array}{cl} \delta_{a,i} & i\in[1,4]\ , \\
c_{a,i} & i>4\ .
\end{array}\right.
\]
for $i=1,2,3,4$. If necessary, we can relabel the $\mathbb{Y}_i$ such that the above holds true. For $a\in[1,4]$, define
\begin{equation}
\mathbb{Y}_a' = \sum_{b=1}^4 M_{ab}\ \Big(\nabla_b \mathbb{X} - (\alpha_b J + \beta_b) \mathbb{X}\Big)\ .
\end{equation}
Then, one has
\begin{equation} \label{Yprime}
\mathbb{Y}_a' = \mathbb{Y}_a + \sum_{j=5}^{n-1} c_{a,j} \mathbb{Y}_j\ .
\end{equation}
where $c_{a,j}$ are some constants. Note that the $q$-series for the $\mathbb{Y}_i$ is not known but the one for $\mathbb{Y}_a'$ are known given $\mathbb{X}$.

When the rank is six, we need to consider a new operator i.e.,
\begin{equation}
\nabla_5 \mathbb{X} - (\alpha_5 J^2 + \beta_5 J + \delta_5) \mathbb{X} = \sum_{i=1}^{n-1}(\gamma_{5,i}\,( J-744)+ \widehat{\gamma}_{5,i}) \, \mathbb{Y}_i \ ,
\end{equation}
The constants $(\alpha_5,\beta_5,\gamma_{5,i})$ are fixed by the leading terms in the $q$-expansion. The sub-leading terms fix the remaining constants $(\delta_5,\widehat{\gamma}_{5,i})$. From practical considerations, we rewrite  the above formula in terms of the known $\mathbb{Y}_a'$.
\begin{equation}
\begin{split}
\nabla_5 \mathbb{X} - (\alpha_5 J^2 + \beta_5 J + \delta_5) \mathbb{X} =& \sum_{a=1}^{4}(\gamma_{5,a}\,( J-744)+ \widehat{\gamma}_{5,a}) \, \mathbb{Y}_a' \\
&+ \sum_{j=5}^{n-1}(\gamma_{5,j}'\,( J-744)+ \widehat{\gamma}_{5,j}') \, \mathbb{Y}_j \ ,
\end{split}
\end{equation}
where $\gamma_{5,j}'=\gamma_{5,j} - \sum_{a=1}^4 \gamma_{5,a}c_{a,j}$ and $\widehat{\gamma}_{5,j}'=\widehat{\gamma}_{5,j} - \sum_{a=1}^4 \widehat{\gamma}_{5,a}c_{a,j}$. For the rank six case, we are done as we can determine $\mathbb{Y}_5$ using the above formula. The remaining $\mathbb{Y}_i$ can now be determined by using Eq. \eqref{Yprime}. 

In the rank six example that we discuss later, it happens that $\gamma_{5,5}'=0$ and one has a straightforward expression for $\mathbb{Y}_5$ which completely determines $\mathbb{Y}_a$ using Eq. \eqref{Yprime}. For the rank six case, it might be simpler to use the operator $\nabla_6 = \frac{1}{\Delta}\mathcal{D}^6$ that was also introduced by Bantay. However, it does not help us determine $\mathbb{Y}_5$ in our example!

\subsection{A $(5,0)$ example}

The affine Lie algebra $F_4$ at level two has five characters and has Wronskian index $\ell=0$.
The central charge is $c=\frac{104}{11}$ and the non-zero conformal dimensions are $h_1 = \frac{6}{11}, h_2 = \frac{9}{11}, h_3=\frac{12}{11}, h_4 = \frac{13}{11}$. The VVMF for the characters has the following $q$-series that we generated using the package SageMath\cite{SageMath}.
\bea
\mathbb{X}(\tau) = \left(\begin{array}{l} q^{-\frac{13}{33}}\left(1 + 52 q + 1430 q^2 + 15184 q^3 + 113412 q^4 + 664248 q^5  +\ldots\right)\\ q^{ \frac{5}{33}} \left(26 + 1352 q + 19279 q^2 + 165556 q^3 + 1070264 q^4 +  \ldots\right)\\ q^{\frac{14}{33}} \left(52 + 1651 q + 19604 q^2 + 154076 q^3 + 937456 q^4 + 4780139 q^5+\ldots\right)\\q^{\frac{23}{33}} \left(273 + 5772 q + 59748 q^2 + 433316 q^3 + 2497313 q^4 + \ldots\right)\\q^{\frac{26}{33}} \left(324 + 6019 q + 60268 q^2 + 425360 q^3 + 2412332 q^4 + \ldots\right)\end{array}\right) \label{5chacsEx2}
\eea
The characters satisfy the $(5, 0)$ MLDE 
\bea
\left(\mathcal{D}^5 - \frac{125}{396} E_4\, \mathcal{D}^3 + \frac{2095}{8712} E_6\, \mathcal{D}^2 - \frac{151375}{1724976} E_4^2 \,\mathcal{D} + \frac{544180}{39135393} E_4 E_6\right) \chi(\tau)=0\ .
\eea

The four new solutions are given by  
\begin{small}
\begin{align*} 
 \mathbb{Y}_1(\tau) &= \left(\begin{array}{l} q^{ \frac{20}{33}}\left(4641 + 649264 q + 25233712 q^2+\ldots\right)\\ q^{ -\frac{28}{33}} \left(1 + 112 q + 66752 q^2 + \ldots\right)\\ q^{\frac{14}{33}} \left(408 + 85680 q + 3902724 q^2+\ldots\right)\\q^{\frac{23}{33}} \left(-4368 - 513084 q - 18561984 q^2+\ldots\right)\\q^{\frac{26}{33}} \left(-12376 - 1237600 q - 41871578 q^2+\ldots\right)\end{array}\right), 
 \end{align*}
 \begin{align*}
\mathbb{Y}_2(\tau) &= \left(\begin{array}{l} q^{ \frac{20}{33}}\left(456 + 24149 q + 462232 q^2+\ldots\right)\\ q^{  \frac{5}{33}} \left(19 + 2488 q + 68096 q^2+ \ldots\right)\\ q^{-\frac{19}{33}} \left(1 - 304 q - 21888 q^2 +\ldots\right)\\q^{\frac{23}{33}} \left(-988 - 45448 q - 822016 q^2+\ldots\right)\\q^{\frac{26}{33}} \left(1976 + 79496 q + 1365112 q^2+\ldots\right)\end{array}\right) \ ,\ 
\mathbb{Y}_3(\tau) = \left(\begin{array}{l} q^{ \frac{20}{33}}\left(84 + 1240 q + 9570 q^2+\ldots\right)\\ q^{ \frac{5}{33}} \left(-8 - 245 q - 2360 q^2+ \ldots\right)\\ q^{\frac{14}{33}} \left(-35 - 680 q - 5628 q^2+\ldots\right)\\q^{-\frac{10}{33}} \left(1 + 120 q + 1660 q^2 +\ldots\right)\\q^{\frac{26}{33}} \left(-50 - 560 q - 4180 q^2+\ldots\right)\end{array}\right), 
\end{align*}
\end{small}
\begin{align*}
\mathbb{Y}_4(\tau) = \left(\begin{array}{l} q^{ \frac{20}{33}}\left(28 + 217 q + 1176 q^2+\ldots\right)\\ q^{ \frac{5}{33}} \left(-7 - 112 q - 686 q^2+ \ldots\right)\\ q^{\frac{14}{33}} \left(20 + 210 q + 1148 q^2+\ldots\right)\\q^{\frac{23}{33}} \left(-14 - 92 q - 511 q^2+\ldots\right)\\q^{-\frac{7}{33}} \left(1 + 28 q + 224 q^2 + \ldots\right)\end{array}\right)
 \end{align*}  
   
The solutions $\mathbb{Y}_1(\tau)$, $\mathbb{Y}_3(\tau)$, and $\mathbb{Y}_4(\tau)$ are admissible characters, once we change an overall negative sign of two of the characters. These characters are also listed in \cite{Duan:2022ltz}. 
We also find new admissible solutions of type $\mathbb{U}_1$ and $\mathbb{U}_2$ (as defined in Eq. \eqref{Udef}) that we list below.
\begin{enumerate}
\item $(\mathbb{X} + \mathbb{Y}_3)$ and $(\mathbb{X} + b\mathbb{Y}_4)$ with $b=1,2,3$ all have Wronskian index $6$
\item  $(\mathbb{X} + \mathbb{Y}_3 + b\mathbb{Y}_4)$ with $b=1,2$ and $(\mathbb{X} + 2\mathbb{Y}_3 + \mathbb{Y}_4)$ all have Wronskian index $12$.
\end{enumerate}

We can read off the values of $r_0^{(i)}$ from the $q$-series for $\mathbb{X}$, We obtain $r_0=(26,52,273,324)$.  In Table \ref{tab4}, we list the new admissible characters that we obtain from the procedure described earlier. In the table, we highlight two cases that appear at special values.

\begin{table}[h] \centering
\begin{tabular}{c|c|c}
\textbf{Admissible Character} & \textbf{Condition} & $\ell$ \\ \hline
$\mathbb{W}_4 = (J+b)\mathbb{X}-\sum_{i=1}^4 r_0^{(i)}\mathbb{Y}_i$ & $b\geq -796$  & $6$ \\ \rowcolor{mColor2}
$\mathbb{W}^{(4)}_3 = \mathbb{W}_4 + r_0^{(4)} \mathbb{Y}_4$ & $b= -730$  & $6$ \\\hline
$\mathbb{W}^{(4)}_3 = \mathbb{W}_4 + r_0^{(4)} \mathbb{Y}_4$  & $b>-730$  & $12$ \\
$\mathbb{W}^{(3)}_3 = \mathbb{W}_4 + r_0^{(3)} \mathbb{Y}_3 $ & $b\geq -733$& 12 \\
$\mathbb{W}^{(2)}_3 = \mathbb{W}_4 + r_0^{(2)} \mathbb{Y}_2 $ & $b\geq -630$& 12 \\
$\mathbb{W}^{(1)}_3 = \mathbb{W}_4 + r_0^{(1)} \mathbb{Y}_2 $ & $b\geq -459$& 12 \\ \rowcolor{mColor2}
$\mathbb{W}^{(12)}_2 = (J+b)\mathbb{X}- r_0^{(1)}\mathbb{Y}_1-r_0^{(2)}\mathbb{Y}_2$ & $b= -646$  & $12$ \\ \hline
$\mathbb{W}^{(12)}_2 = (J+b)\mathbb{X}- r_0^{(1)}\mathbb{Y}_1-r_0^{(2)}\mathbb{Y}_2$ & $b\geq-646$  & $18$ \\ 
$\mathbb{W}^{(13)}_2 = (J+b)\mathbb{X}-r_0^{(1)} \mathbb{Y}_1-r_0^{(3)}\mathbb{Y}_3$ & $b>-755$  & $18$ \\
$\mathbb{W}^{(14)}_2 = (J+b)\mathbb{X}- r_0^{(1)}\mathbb{Y}_1-r_0^{(4)}\mathbb{Y}_4$ & $b\geq-447$  & $18$ \\
$\mathbb{W}^{(23)}_2 = (J+b)\mathbb{X}- r_0^{(2)}\mathbb{Y}_2-r_0^{(3)}\mathbb{Y}_3$ & $b\geq-487$  & $18$ \\
$\mathbb{W}^{(24)}_2 = (J+b)\mathbb{X}- r_0^{(2)}\mathbb{Y}_2-r_0^{(4)}\mathbb{Y}_4$ & $b\geq-417$  & $18$\\
$\mathbb{W}^{(34)}_2 = (J+b)\mathbb{X}- r_0^{(3)}\mathbb{Y}_3-r_0^{(4)}\mathbb{Y}_4$ & $b\geq-661$  & $18$ \\ \hline
$\mathbb{W}^{(1)}_1 = (J+b)\mathbb{X}-r_0^{(1)} \mathbb{Y}_1$ & $b\geq-571$  & $24$ \\
$\mathbb{W}^{(2)}_1 = (J+b)\mathbb{X}-r_0^{(2)} \mathbb{Y}_2$ & $b\geq-445$  & $24$ \\
$\mathbb{W}^{(3)}_1 = (J+b)\mathbb{X}- r_0^{(3)}\mathbb{Y}_3$ & $b\geq-645$  & $24$ \\
$\mathbb{W}^{(4)}_1 = (J+b)\mathbb{X}-r_0^{(4)} \mathbb{Y}_4$ & $b\geq-651$  & $24$ \\ \hline

\end{tabular}
\caption{Admissible characters with $c=\frac{368}{11}$ obtained using the new solutions}\label{tab4}
\end{table}

The solutions that we obtain are solutions to the Matrix MLDE given in Eq. \eqref{MMDE} with 
\begin{align*}
\Lambda &= \text{Diag}\left(
 -\frac{13}{33} , -\frac{28}{33} , -\frac{19}{33},-\frac{10}{33}, -\frac{7}{33} 
\right)\ , \\
\mathcal{Y} &=\begin{pmatrix}
52 & 4641 & 456 & 84 & 28 \\
 26 & 112 & 19 & -8 & -7 \\
 52 & 408 & -304 & -35 & 20 \\
 273 & -4368 & -988 & 120 & -14 \\
 324 & -12376 & 1976 & -50 & 28
\end{pmatrix}\ .
\end{align*}

 \subsection{A rank six example}
 
The tricritical Ising model is  the Virasoro  minimal model $\mathcal{M}(5,4)$ with central charge $c=\tfrac7{10}$ and  six primary fields. The non-zero conformal dimensions are  $\{\tfrac{7}{16}, \tfrac1{10}, \tfrac35, \tfrac3{80}, \tfrac32\}$. The characters  are given by
\begin{eqnarray}
\mathbb{X} = \left(\begin{array}{l} 
q^{-\frac{7}{240}}\left(1 + q^2 + q^3 + 2 q^4 + 2 q^5 + 4 q^6   +\ldots\right)\\ 
q^{ \frac{49}{120}} \left(1 + q + q^2 + 2 q^3 + 3 q^4 + 4 q^5 + \ldots\right)\\ 
q^{\frac{17}{240}} \left(1 + q + q^2 + 2 q^3 + 3 q^4 + 4 q^5+\ldots\right)\\
q^{\frac{137}{240}} \left(1 + q + 2 q^2 + 2 q^3 + 4 q^4 + 5 q^5+\ldots\right)\\
q^{\frac{1}{120}} \left(1 + q + 2 q^2 + 3 q^3 + 4 q^4 + 6 q^5+\ldots\right) \\
q^{\frac{113}{240}} \left(0+q + q^2 + 2 q^3 + 2 q^4 + 3 q^5 + 4 q^6+\ldots\right) 
\end{array}\right) \label{TIMchacs} 
\end{eqnarray}
The characters satisfy a $(6, 0)$ MLDE
\be
\begin{split}
\left(\mathcal{D}^6 -  \frac{16331}{28800}   E_4\, \mathcal{D}^4 +   \frac{36181}{691200}     E_6\, \mathcal{D}^3 +   \frac{7436473}{3317760000}     E_4^2\, \mathcal{D}^2 \right.\hspace{1.3in}\\ - \frac{29669933}{13271040000}    E_4 E_6\, \mathcal{D} 
\left.+  \frac{108360091}{1990656000000}    E_4^3 - \frac{115305407}{1911029760000}    E_6^2\right) \chi(\tau)=0 \ .
\end{split}\nonumber
\ee
However, we write the exponent in the last line as if were a solution to a $(6,6)$ MLDE. This is because all the other solutions that we generate turn out to have Wronskian index $6$. In a sense, the zero that we write out explicitly in the last line of the vvmf is an accidental one. One has
\[
\Lambda = \text{Diag}\left(-\tfrac{7}{240}, -\tfrac{71}{120}, -\tfrac{223}{240}, -\tfrac{103}{240}, -\tfrac{119}{120}, -\tfrac{127}{240}\right)\ ,
\]
and the exponents of $\mathbb{X}$ then given by \eqref{lambdalpharel}.

We then find that remaining five solutions, all with Wronskian index $6$, given by
\begin{eqnarray*}
\begin{small}
\mathbb{Y}_1 = \left(\begin{array}{l} q^{ \frac{233}{240}}\left(3104   + 105344 q  +\ldots\right)\\ q^{ -\frac{71}{120}} \left(1  + 11 q  - 87 q^2+ \ldots\right)\\ q^{\frac{17}{240}} \left(-32   - 7520 q +\ldots\right)\\q^{\frac{137}{240}} \left(800   + 46976 q +\ldots\right)\\q^{\frac{1}{120}} \left(-6   + 33 q +\ldots\right) \\q^{\frac{113}{240}} \left(-288  - 20576 q +\ldots\right) \end{array}\right)\ ,\ 
\mathbb{Y}_2 = \left(\begin{array}{l} q^{ \frac{233}{240}}\left(44625   + 4213110 q +\ldots\right)\\ q^{  \frac{49}{120}} \left(-1785  - 550137 q + \ldots\right)\\ q^{-\frac{223}{240}} \left(1  - 41 q  - 46498 q^2+\ldots\right)\\q^{\frac{137}{240}} \left(-2500   - 497227 q +\ldots\right)\\q^{\frac{1}{120}} \left(7  + 47623 q +\ldots\right) \\q^{\frac{113}{240}} \left(1989  + 516783 q +\ldots\right) \end{array}\right), 
 \end{small} 
\end{eqnarray*}
\begin{eqnarray*}
\begin{small}
\mathbb{Y}_3 = \left(\begin{array}{l} q^{ \frac{233}{240}}\left(799   + 14756 q +\ldots\right)\\ q^{  \frac{49}{120}} \left(119   + 4471 q + \ldots\right)\\ q^{ \frac{17}{240}} \left(-10   - 689 q +\ldots\right)\\q^{-\frac{103}{240}} \left(1  - 98 q  - 3163 q^2 +\ldots\right)\\q^{\frac{1}{120}} \left(-9   - 777 q +\ldots\right) \\q^{\frac{113}{240}} \left(119   + 3791 q +\ldots\right) \end{array}\right)\ , \ 
\mathbb{Y}_4 = \left(\begin{array}{l} q^{ \frac{233}{240}}\left(95200   + 10638464 q +\ldots\right)\\ q^{  \frac{49}{120}} \left(-34   + 340 q + \ldots\right)\\ q^{ \frac{17}{240}} \left(32   + 101728 q +\ldots\right)\\q^{ \frac{137}{240}} \left(-4896   - 1181568 q +\ldots\right)\\q^{-\frac{119}{120}} \left(1  - 17 q  + 221 q^2 +\ldots\right) \\q^{\frac{113}{240}} \left(-3808  - 1208224 q +\ldots\right) \end{array}\right), 
\end{small} 
\end{eqnarray*}
\begin{eqnarray*}
\mathbb{Y}_5 = \left(\begin{array}{l} q^{ \frac{233}{240}}\left(1250   + 34673 q  + 475726 q^2 + 4495681 q^3 + 33177533 q^4  +\ldots\right)\\ q^{  \frac{49}{120}} \left(-143   - 8591 q  - 163471 q^2 - 1864478 q^3 - 15636141 q^4 + \ldots\right)\\ q^{ \frac{17}{240}} \left(11   + 3014 q  + 74422 q^2 + 
 977725 q^3  +\ldots\right)\\q^{ \frac{137}{240}} \left(375   + 16690 q  + 285448 q^2 + 3067292 q^3 + 24729901 q^4 +\ldots\right)\\q^{ \frac{1}{120}} \left(-15  - 3343 q  - 88222 q^2 - 1191341 q^3 - 11130300 q^4 +\ldots\right) \\q^{-\frac{127}{240}} \left(1  + 157 q  + 7412 q^2 + 
 136386 q^3 + 1516832 q^4 + \ldots\right) \end{array}\right). 
\end{eqnarray*} 
The solution $\mathbb{Y}_5(\tau)$ is admissible characters, once we change an overall negative sign of two of the characters.  We find \textit{no} new admissible characters from the $\mathbb{U}_r$.

We can read off the values of $r_0^{(i)}$ from the $q$-series for $\mathbb{X}$, We obtain $r_0=(1,1,1,1,0)$. The zero in the last entry implies that $\mathbb{Y}_5$ cannot appear. Thus, only four of the five new characters appear in our new family of admissible characters.  In Table \ref{tab5}, we list the new admissible characters that we obtain from the procedure described earlier. In the table, we highlight two cases that appear at special values. Observe that the set of admissible characters that we obtain this way are way smaller than seen in Table \ref{tab4} in the five character example. It also emphasises the fact that this method is not guaranteed to lead to admissible characters in general. The surprise is that it does.

\begin{table}[h] \centering
\begin{tabular}{c|c|c}
\textbf{Admissible Character} & \textbf{Condition} & $\ell$ \\ \hline
$\mathbb{W}_4 = (J+b)\mathbb{X}-\sum_{i=1}^4 \mathbb{Y}_i$ & $b\geq -744$  & $12$ \\ \hline \rowcolor{mColor2}
$\mathbb{W}^{(2)}_3 = \mathbb{W}_4 + \mathbb{Y}_2 $ & $b=-649$& 12 \\
$\mathbb{W}^{(2)}_3 = \mathbb{W}_4 + \mathbb{Y}_2 $ & $b> -649$& 18 \\
$\mathbb{W}^{(3)}_3 = \mathbb{W}_4 + \mathbb{Y}_3 $ & $b\geq -744$& 18 \\
$\mathbb{W}^{(1)}_3 = \mathbb{W}_4 + \mathbb{Y}_1 $ & $b\geq -744$& 18 \\ \hline \rowcolor{mColor2}
$\mathbb{W}^{(13)}_2 = (J+b)\mathbb{X}- \mathbb{Y}_1-\mathbb{Y}_3$ & $b=-43$  & $18$ \\
$\mathbb{W}^{(13)}_2 = (J+b)\mathbb{X}- \mathbb{Y}_1-\mathbb{Y}_3$ & $b>-43$  & $24$ \\
$\mathbb{W}^{(14)}_2 = (J+b)\mathbb{X}- \mathbb{Y}_1-\mathbb{Y}_4$ & $b\geq-744$  & $24$ \\
$\mathbb{W}^{(24)}_2 = (J+b)\mathbb{X}- \mathbb{Y}_2-\mathbb{Y}_4$ & $b\geq-744$  & $24$\\
$\mathbb{W}^{(34)}_2 = (J+b)\mathbb{X}- \mathbb{Y}_3-\mathbb{Y}_4$ & $b\geq-660$  & $24$ \\ \hline
$\mathbb{W}^{(1)}_1 = (J+b)\mathbb{X}- \mathbb{Y}_1$ & $b\geq55$  & $30$ \\
 \hline
 $\mathbb{W}^{(4)}_1 = (J+b)\mathbb{X}- \mathbb{Y}_4$ & $b\geq-713$  & $30$ \\
 \hline
\end{tabular}
\caption{Admissible characters with $c=\frac{247}{10}$ obtained using the new solutions}\label{tab5}
\end{table}
We conclude by giving the matrix MLDE whose solutions are these six characters. The matrix of exponents $\Lambda$ have already been given above. The matrix $\mathcal{Y}$ is given by
\begin{equation}
\mathcal{Y}= 
\begin{pmatrix}
 0 & 3104 & 44625 & 799 & 95200 & 1250 \\
 1 & 11 & -1785 & 119 & -34 & -143 \\
 1 & -32 & -41 & -10 & 32 & 11 \\
 1 & 800 & -2500 & -98 & -4896 & 375 \\
 1 & -6 & 7 & -9 & -17 & -15 \\
 0 & -288 & 1989 & 119 & -3808 & 157 
\end{pmatrix}\ .
\end{equation}
Solving the Matrix MLDE using the Frobenius method is one way to simultaneously generate all six solutions.

\section{Conclusion}

In this paper, we have seen how to use the VVMF approach to provide a complement to the MMS approach to classifying RCFTs. The main ingredient is that one has to start with a known RCFT or at least an admissible solution, that we called $\mathbb{X}$, generated using the MMS approach. Then the VVMF method lets one generate new solutions (denoted by $\mathbb{Y}_i$, $i=1,\ldots \textrm{rank}-1)$) that share the same $S$ and $T$ matrix as the original solution. We then show how to construct potentially new admissible solutions by taking particular linear combinations of $(aJ+b)\mathbb{X}$ and the new solutions. The linear combination was restricted by the requirement that the Wronskian index takes minimal value. This left us with only one parameter $b$ to fix. The following generalisations can be carried out.
\begin{enumerate}
\item Relax the condition on the Wronskian index and study the conditions on the larger parameter space given by the coefficients of the new solutions. This could, in principle, give rise to a larger family of solutions.
\item Let $P_m(x)$ be a monic polynomial of degree $m$ in $x$. For $m>1$, start with $P_m(J) \mathbb{X}$ and look for new admissible solutions by adding linear combinations of the new solutions. All such solutions have central charge $(c+24m)$, where $c$ is the central charge of the original RCFT.
\end{enumerate}

The work of Kaidi et al.\cite{Kaidi:2021ent}, provides a list of exponents (modulo one) for quasi-characters with up to five characters. This provides  the entries in the matrix of exponents $\Lambda\mod 1$. It would be nice to combine their results with the Matrix MLDE to come up with a new approach the holomorphic modular bootstrap. Knowing the $S$-matrix would be useful as it provides conditions on the eigenvalues of $S$ and $U=ST^{-1}$ through Eq. \eqref{RiRoc1}. The matrix $\mathcal{Y}$ will then be determined by self-consistency. As a first step, one can systematically look for admissible solutions starting from linear combinations known solutions and quasi-characters obtained using the VVMF approach. This is being currently studied by us\cite{ongoing}.

The work on the classification of Modular Tensor Categories(MTC) up to rank four was carried out in \cite{Rowell:2009}. This work also discusses formulae for the $S$-matrices among other things.The $S$-matrix for them $(3,3)$ model appears in their list of rank three examples and is realised by an MTC that they call $(A_1,5)_{\tfrac12}$.  Gepner and Kapustin also obtain this model in their classification of fusion rules for up to six primaries\cite{Gepner:1994bb}. In Table 3, this potential RCFT is listed as a $\BZ_2$ orbifold of the $A_1$ affine Kac-Moody Lie algebra at level $5$. This presumably indicates a restriction to $\BZ_2$ invariant primaries of the $SU(2)_5$ affine Kac-Moody Lie algebra.
The VVMF approach to RCFT might also provide inputs to the classification of MTCs. This is worth studying. 

\medskip
\noindent \textbf{Acknowledgments:} JS is indebted to Chethan N. Gowdigere, from whom he learned the subject. We would like to thank Sunil Mukhi, Arpit Das, Sachin Kala, Bobby Ezhuthachan, Dileep Jatkar, Rajath Radhakrishnan, Akhila Sadanandan, Nemani V. Suryanarayana, Koushik Ray, and Sudipta Das for their helpful discussions. Both authors wish to thank the organisers of the  School and Workshop on Number Theory and Physics  held at  Abdus Salam International Center for Theoretical Physics at Trieste during the summer of 2024 where some of this work was done. JS is supported by an Instiute Postdoctoral Fellowship at IIT Madras.

\clearpage

\appendix

\section{Modular Forms}

In this appendix we define some modular forms of $PSL(2,\BZ)$ appear in this paper. This is to provide a quick description to set up notations. For details, refer to the introduction to the topic by Zagier\cite{Zagier:1992,Zagier:2008}. Also see \cite{Dabholkar:2012}.

A modular form of weight $w$  is a holomorphic function on the Poincar\'e upper half plane $\mathbb{H}$ such that
\begin{equation}
\phi\left(\tfrac{a\tau+b}{c\tau+d}\right) = (c\tau+d)^w\  
\phi(\tau)
\end{equation}
for $\left(\begin{smallmatrix} a & b \\  c & d \end{smallmatrix}\right)\in PSL(2,\BZ)$. The invariance under $\tau\rightarrow \tau+1$ implies that one has the Fourier series
\[
\phi(\tau) = \sum_{n=-\infty}^{\infty} a_n q^n\ .
\]
If the modular form is bounded as $\tau \rightarrow i\infty$ (or $q\rightarrow0$), one has $a_n=0$ for $n<0$. Such modular forms are called \textit{holomorphic} modular forms. A modular form is called weakly holomorphic if $a_n=0$ for some $n<-N$ where $N$ is a positive integer. 

 We  provide explicit formulae for the Eisenstein series $E_2, E_4,$ and $E_6$, the Dedekind-$\eta$ function, the cusp form of weight  $12$, and Klein-$J$ invariant. We define $q=\exp(2\pi i \tau)$. 
\begin{align*}
E_2(\tau) &= 1 - 24\sum\limits_{n=1}^{\infty} \frac{n\, q^n}{(1-q^n)} = 1-24q-72q^2-96q^3  +\ldots \\
E_4(\tau) &= 1 + 240\sum\limits_{n=1}^{\infty} \frac{n^3\, q^n}{(1-q^n)} = 1 + 240 q + 2160 q^2 + 6720 q^3  + \ldots \\
E_6(\tau) &= 1 - 504\sum\limits_{n=1}^{\infty} \frac{n^5\, q^n}{(1-q^n)} = 1 - 504 q - 16632 q^2 - 122976 q^3 +\ldots \\
\eta(\tau) &= q^{\frac{1}{24}}\left(1+\sum\limits_{n=1}^{\infty} (-1)^n(q^{\frac{n(3n-1)}{2}} + q^{\frac{n(3n+1)}{2}})\right) \\ &= q^{\frac{1}{24}}(1 - q - q^2 + q^5 + \ldots) \\
\Delta(\tau) &= \eta^{24} =  \frac{E_4^3 - E_6^2}{1728} = q - 24 q^2 + 252 q^3 - 1472 q^4 + 4830 q^5 + \ldots \\
J(\tau) &= \frac{E_4^3}{\Delta} = \frac{1}{q} + 744 + 196884 q + 21493760 q^2 +\ldots
\end{align*}
Of these, only the Dedekind eta function has weight half and is a modular form of a sub-group of $PSL(2,\BZ)$ as it transforms with phases that are 24-th roots of unity. Further, only the $J$-invariant is a weakly holomorphic modular function while the others are holomorphic modular forms. 

\noindent The Serre-Ramanujan covariant derivative, acting on a modular form of weight $w$,  is defined as
\bea
\mathcal{D}_w := q\frac{d}{dq} - \frac{w}{12} E_2.
\eea
It maps a modular form of weight $w$ to one of weight $(w+2)$. Thus higher powers of the covariant derivative need to take this into account. One has
\bea
\mathcal{D}_w^n = \left(q\frac{d}{dq} - \frac{w+2n-2}{12} E_2\right)\circ  \mathcal{D}_w^{n-1}\ .
\eea
Quite often, we do not specify the weight of the modular form as it is clear in context. We write $\mathcal{D}$ in place of $\mathcal{D}_w$.

\end{document}